\documentclass[11pt,twoside]{article}

\usepackage{asp2014}

\aspSuppressVolSlug
\resetcounters

\begin{document}

\title{Extracting Insights from Astrophysics Simulations}

\author{Nathan~J.~Goldbaum$^1$}
\affil{$^1$National Center for Supercomputing Applications, University of
  Illinois at Urbana-Champaign, Urbana, IL, USA; \email{ngoldbau@illinois.edu}}

\paperauthor{Nathan~J.~Goldbaum}{ngoldbau@illinois.edu}{0000-0001-5557-267X}{University
of Illinois at Urbana-Champaign}{National Center for Supercomputing Applications}{Urbana}{IL}{61801}{USA}

\begin{abstract}
Simulations inform all aspects of modern astrophysical research, ranging in scale from 1D and 2D test problems that can run in seconds on an astronomer's laptop all the way to large-scale 3D calculations that run on the largest supercomputers, with a spectrum of data sizes and shapes filling the landscape between these two extremes. I review the diversity of astrophysics simulation data formats commonly in use by researchers, providing an overview of the most common simulation techniques, including pure N-body dynamics, smoothed particle hydrodynamics (SPH), adaptive mesh refinement (AMR), and unstructured meshes. Additionally, I highlight methods for incorporating physical phenomena that are important for astrophysics, including chemistry, magnetic fields, radiative transport, and "subgrid" recipes for important physics that cannot be directly resolved in a simulation. In addition to the numerical techniques, I also discuss the communities that have developed around these simulation codes and argue that increasing use and availability of open community codes has dramatically lowered the barrier to entry for novice simulators.
\end{abstract}

\section{Introduction}
Direct 3D numerical simulations of astrophysical phenomena are a key driver of theoretical prediction in 21st century astrophysics. By offering a dynamical view of processes that are unobservable on human timescales, numerical simulations allow researchers to extract unique insights that are unavailable from other approaches due to the complexity of the underlying physical systems. Due to the nature of the numerical discretizations driving the simulation codes, increasing computer power will naturally bring more and more phenomena within the envelope of computational feasability. As increased power is coupled with improved methods and modeling techniques, astrophysical prediction is increasingly informed by direct numerical simulations.

It is therefore not surprising that the astrophysics simulation software ecosystem is quite diverse. Different systems lend themselves to different simulation techniques, results are validated by being reproduced from independent codebases, and individual research groups are incentivized to develop software both to produce new results and in some cases to make proprietary use of new technology to produce novel research results. Partially as a reaction to these incentives, open research communities are developing around astrophysics simulation codes and inclreasingly simulation software is made publicly available so that it can be used and reviewed by anyone.

Here I will review the ecosystem of astrophysical simulation software and simulation techniques. This will by no means be a comprehensive review. Instead, I will primarily focus on generally available community research software. I will also mention several propietary research codes that have made a substantial impact on the research landscape in terms of published results. I will also attempt to focus on codes that are still used to generate research results, although I will mention a few older codes for historical context. The review will be organized according to the simulation technique used by the software. First, in Section~\ref{nbody}, I will discuss pure N-body and smoothed particle hydrodynamics (SPH) codes. Next, in Section~\ref{amr}, I will discuss grid-based hydrodynamics codes, with a focus on adaptive mesh refinement (AMR) codes. Finally, in Section~\ref{usm}, I will discuss codes that make use of novel computational techniques developed only recently.

\section{N-body Methods}
\label{nbody}

In this approach, an astrophysical system is discretized into a system of self-gravitating particles which interact via their mutual gravitational attraction. In most cases the particles are simply test particles which are used to sample the gravitational force of the system, but each individual particle does not necessarily have any physical meaning. N-body methods are Lagrangian and thus well-suited for astrophysical problems that tend to have large dynamic ranges in the sizes of structures. Rather than wasting computation on regions in which not much is happening, work naturally flows to where the mass is located. N-body methods in astrophysics can be further subdivided into pure N-body methods, which only simulate the gravitational force, and SPH methods, which can include the effect of gas dynamics. We describe these two approaches below.

Using pure N-body techniques to understand physical systems has a rich history \citep[see e.g.\ ][]{holmberg1941, hoerner1960, peebles1970, press1974}. In some cases, such as simulations of star clusters \citep{wang2016}, or even galaxies \citep{bedorf2014}, the masses of particles in the simulation approach the masses of the physical bodies that make up the real gravitational system, however this is usually unattainable.

Since gravity is a long-range force, this means that the gravitational force calculation for any given particle depends on the masses and positions of every other particle in the simulation. The worst-case naive algorithm that evaluates the pairwise gravitational force for all particles is $O(N^2)$. State-of-the-art N-body calculations include as many as a trillion particles and are run on clusters containing thousands of compute nodes. The naive pairwise algorithm would simply not be feasible to run on simulations of this size. Instead, N-body codes make use of some form of spatial acceleration to convert the calculation of the gravitational force into an $O(N \log N)$ calculation. Often, this involves depositing the N-body data onto a mesh and then making use of a fast Fourier transform (FFT) to solve the Poisson equation on the mesh, this is known as a Particle-mesh (PM) algorithm \citep{hockney1988}. Another alternative is to calculate the gravitational acceleration by making use of an octree data structure. Particles contained in octree zones that are sufficiently far away (usually quantified in terms of an opening angle relative to the particle under consideration) are grouped together, and the collective effect of distant particles is averaged, possibly using a multipole expansion. This algorithm is originally due to \citet{barnes1986}, and octrees used for a gravitational force calculation are known as a Barnes-Hut tree. These two approaches can be combined to produce a TreePM code \citep{bagla2002}, with a tree used to calculate the influence of nearby particles, and mesh deposition used for the forces from more distant particles. This allows much easier implementation of periodic boundary conditions useful for cosmology simulations while retaining the high force resolution of the Barnes-Hut algorithm.

Many problems include the coupled effects of particle dynamics and gas dynamics, so it becomes very appealing to directly include gas dynamics in an N-body model. \citet{lucy1977} and \citet{gingold1977} introduced the concept of smoothed particle hydrodynamics, where N-body particles are used as Lagrangian markers to sample an underlying continuous field. Rather than defining the properties of the gas based on local properties of any given particle, the properties of the gas are calculated based on a weighted average of the value of a field at the locations of nearby particles. The weighting function, known as the smoothing kernel, defines a finite compact region where particles may interact with each other. After calculating the gravitational forces (if any) using the N-body algorithms described above, the hydrodynamic forces may be calculated by solving a discretized version of the equiations of fluid dynamics, where the individual terms in the equations are calcualted by averaging over a smoothing kernel. In recent years, SPH algorithms have come under criticism for not being able to resolve certain fluid instabilities \citep{agertz2007} and artificially suppressing mixing \citep{read2010}. This has led to the development of alternative SPH formalisms \citep{abel2011, hopkins2013}, as well as moving mesh and advanced meshless methods (see Section~\ref{usm} for more discussion and see \citet{price2012} for a detailed recent review of modern SPH formalism).

\subsection{N-body and SPH Simulation Codes}

The \texttt{Gadget} family of SPH codes is quite possibly the most commonly used astrophysical simulation code. Originally created in 1998 by Volker Springel as part of his PhD project, \texttt{Gadget-1} was released publicly in 2000 \citep{springel2001}. Later, the next-generation \texttt{Gadget-2} code was released, featuring a complete rewrite in C, improved algorithms, including an entropy-conserving approach for forumulating SPH, and a TreePM implementation \citep{springel2005b} that makes \texttt{Gadget-2} suitable for large-scale N-body calculations of structure formation. This capability was used to run the Millenium simulations \citep{springel2005a}, at the time the largest pure N-body simulation of structure formation in the universe, using more than $10^{10}$ particles. \texttt{Gadget-3} was developed afterwards but was never released publicly. It includes an improved domain decomposition algorithm and improvements to the SPH formalism. \texttt{Gadget} has been used extensively to study astrophysical phenomena at many scales, including the impact of gas on cosmological structure formation \citep{schaye2015}, isolated galaxies \citep{springel2005c} and disk galaxy collisions \citep{robertson2006}, the formation of the first stars \citep{clark2011}, and even simulations of the formation of the Moon \citep{jackson2012}.

\texttt{Gadget-2} was released under the terms of the GNU General Public License v2 in 2005. It is available as a tarball at \url{http://wwwmpa.mpa-garching.mpg.de/gadget/}. The public release includes an adiabatic hydrodynamics module, self gravity, and can be used in comoving coordinates. The public release does not include any additional physics modules. This lack of physics modules in the public version has led to the development of many independent private versions of \texttt{Gadget-2} that include alternate implementations of physics capabilities such as cooling and heating processes, star formation and feedback, AGN feedback, and sink particles. Rather than any single \texttt{Gadget} community, there are instead balkanized islands of code development within individual research groups.

\texttt{Tipsy} (\url{https://github.com/N-BodyShop/tipsy}) is a data format and visualization tool originally written by Tom Quinn in the late 80's. The data format is shared by a number of simulation codes whose lineage traces back to the N-body shop research group at the University of Washington. \texttt{PKDGRAV} \citep{stadel2001} is a pure N-body dynamics code. Rather than making use of an octree to calculate the gravitational force, it instead makes use of a parallel $k$-d tree. Compared with an octre, a $k$-d tree has better data balance and can be constructed more efficiently. This property enabled extremely high resolution simulations of galaxy formation in $\Lambda$-CDM cosmology, such as the via lactea simulations \citep{diemand2008}. More recently, \texttt{PKDGRAV} was ported to GPUs \citep{potter2017} and made publicly available (\url{https://bitbucket.org/dpotter/pkdgrav3/}), although without a license declaration.

\texttt{Gasoline} \citep{wadsley2004} (\url{http://gasoline-code.com}) is an SPH code originally built on top of \texttt{PKDGRAV}.\@ Similar to how \texttt{PKDGRAV} was used most impressively to elucidate the details of Milky Way mass dark matter halos, gasoline has been used quite heavily to understand the details of disk galaxy formation using high resolution simulations and improved physical models, including the blastwave feedback model of \citet{stinson2006} and metal line cooling and metal mixing models of \citet{shen2010}. This allowed others to produce advanced galaxy simulations \citep{guedes2011, pontzen2012, zolotov2012}. More recently, the next generation version of the code, \texttt{Gasoline2}, was described in \citet{wadsley2017}. Gasoline2 was recently made publicly available on GitHub (\url{https://github.com/N-BodyShop/gasoline}) under the GPLv2. This version includes an updated SPH formalism. Recently, the physical models available in the \texttt{Gasoline} codebase were incorporated into the \texttt{ChaNGa} code \citep{jetley2008}, a research codebase built on top of the \texttt{Charm++} message passing library. \texttt{Charm++} enables \texttt{ChaNGa} to scale to higher CPU counts and obtain improved efficiency by enabling asynchronous calculation of the various steps in the simulation timestep by the compute nodes. \texttt{ChaNGa} is publicaly available at \url{https://github.com/N-BodyShop/changa} under the GPLv2.

Other public SPH codes include \texttt{SEREN} \citep{hubber2011}  and \texttt{PHANTOM} \citep{price2017}. \texttt{SEREN} is available at \url{http://dhubber.github.io/seren} under the GPLv2. It has been heavily used in simulations of star cluster formation and includes prescriptions for sink particles \citep{walch2013} as well as timestepping schemes suitable for tracking N-body interactions in clusters. \texttt{PHANTOM} is available at \url{https://phantomsph.bitbucket.io} under the GPLv3 and includes an advanced implementation of the SPH formalism, MHD capabilities, and subgrid models useful for studying the ISM and galaxies.

\section{Adaptive Mesh Refinement}
\label{amr}

Eulerian meshes are able to better capture shocks and instabilities compared with traditional SPH codes. In addition, SPH codes have historically had a difficult time incorporating magnetic fields, while this is straightforward in an Eulerian formulation. For this reason, Eulerian codes have also found an active niche in the research landscape.

In an early example of an Eulerian code being used in astrophysics, \citet{boss1979} numerically investigated the 3D collapse of a non-axisymmetric rotating cloud of gas in an attempt to understand the formation of binary stars, making use of a 3D Eulerian mesh with as few as ${\sim}10^4$ zones. Later work using Eulerian methods focused on improving robustness, accuracy, and spatial resolution. In an effort to improve robustness and apply a single code to many different problems \citet{stone1992} introduced the \texttt{Zeus} code. \texttt{Zeus} employed simple algorithms enabling rapid code development and compotational robustness, solving the coupled evolution of gas, magnetic fields, and gravity on a staggered mesh using a finite difference scheme. \texttt{Zeus} can be thought of as an ancestor to modern astrophysics simulation codes in its aim to provide a platform for simulating a wide variety of phenomena, however other codes provided a basis for modern algorithms.

The development of the adaptive mesh refinement (AMR) technique paired with a hydrodynamics scheme with second-order accurate shock-capturing in the 1980's proved fundamental to the development of modern Eulerian codes used in astrophysics. \citet{berger1984} presented the earliest example of structured AMR, where the computation is allowed to proceed at multiple independent resolution levels. In the same year \citet{colella1984} presented the piecewise parabolic method (PPM), demonstrating a shock capturing code with improved accuracy and reduced artificial diffusivity compared with other methods at a fixed computational cost. Finally \citet{berger1989} described a code synthesizing AMR and PPM hydrodynamics. This code is the direct ancestor of many of the modern astrophysics codes described below.

A structured AMR code segments the computational domain into rectilinear subsets, normally referred to as blocks or grids. An individual grid defines a subregion of the computational volume at a fixed resolution and may overlap with grids at higher or lower resolution levels. Usually grids are nested, with the hierarchy of grids defining a tree data structure. The equations of hydrodynamics can be solved independently within each grid with boundary conditions obtained from the state of the computation on neighboring grids patches. The need to fill in boundary conditions in so-called ``ghost'' zones means that AMR methods incur numerical overhead compared with SPH codes, which do not need to track extra fluid elements for bookkeeping purposes, except perhaps across different nodes in a cluster computer. Instead of structured AMR, some codes use a volume-filling octree. This approach allows better memory efficiency while only allowing for first-order accuracy since only a single ghost zone is guaranteed to be available.

\subsection{AMR Codes}

\texttt{Boxlib} \citep{zhang2016} and \texttt{Chombo} \citep{adams2015} are software frameworks for building AMR codes. They handle the bookkeeping for the grid data structure as well as I/O and communication primitives. Individual codes implement solvers for various physics applications. Generally these codes do not aim for general applicability. Instead they are developed by individual research groups for a primary research problem. \texttt{Chombo} was forked from \texttt{Boxlib} in 1998 so these frameworks share many concepts. More recently these efforst have conslidated into the \texttt{AMReX} framework (\url{https://amrex-codes.github.io/}), and future efforts will be in the context of \texttt{AMReX}.

There are a number of codes in the \texttt{Boxlib} and \texttt{Chombo} families. An early result from the \texttt{Orion} code \citep{klein1999, li2012} demonstrated the minimum resolution necessary to resolve the collapse of a self-gravitating cloud of gas  \citep{truelove1998}, a key process in many astrophysics simulations. Later \texttt{Orion} simulations included increasingly realistic physical effects for simulating the collapse of molecular clouds into protostars including non-ideal MHD and radiation transport \citep{myers2014, rosen2016}. \texttt{Orion} is not publicly available. The \texttt{PLUTO} code \citep{mignone2012} is publicly available (\url{http://plutocode.ph.unito.it/}) under the GPLv2 and includes a number of advanced physics packages, including support for MHD, relatavistic hydrodynamics (RHD), dissipative and radiative processes, and physical setups such as shearing boxes and non-cartesian coordinate systems. \texttt{PLUTO} supports static mesh hierarchies so it is particularly well suited for problems like disks that have roughly fixed geometries. Examples include simulations of massive star formation \citep{kuiper2010}, simulations of magnetized disk accretion \citep{lesur2014}, and simulations of gas in the galactic center environment \citep{burkert2012}. \texttt{Nyx} \citep{almgren2013} focuses on scalable large-scale cosmology simulations to aid in constraining observations of cosmological parameters \citep{sorini2016}. It is publicly available (\url{https://github.com/AMReX-Astro/Nyx}) and is publicly developed under the 3-clause BSD license. Finally, the \texttt{Maestro} and \texttt{Castro} codes are primarily used for the detailed modeling progenitors of supernmova explosions. \texttt{Maestro} \citep{nonoka2014} includes solvers tuned for low Mach number flows, while \texttt{Castro} \citep{almgren2010} includes shack capturing and capabilities for resolving highly compressible flows. Both include sophisticated radiation transport and nuclear burning capabilities. Both codes are currently developed in the open on Github under the 3-clause BSD license (see \url{https://github.com/AMReX-Astro/MAESTRO} and \url{https://github.com/AMReX-Astro/Castro}).

\texttt{Enzo} \citep{bryan2014} is another example of a openly developed community astrophysics code. Originally developed by Greg Bryan for his PhD thesis, in subsequent years a variety of research groups worked on private, forked versions of the \texttt{Enzo} code. Later these disparate research groups elected to merge their divergent codebases and publicly release \texttt{Enzo} (\url{http://enzo-project.org}). To this day, \texttt{Enzo} aims to be a community-developed and maintained code, with development happening under a 3-clause BSD license, code review and testing handled by a team of core developers, and new code contributions landing as people develop them for their research. In the \texttt{Enzo} community there is an expectation that newly developed physical models and features should get upstreamed to the public version for shared development. For this reason \texttt{Enzo} supports a wide variety of physical models, including several MHD solvers, radiation, chemistry, as well as star formation and feedback models. Examples of major science results produced using the \texttt{Enzo} code include simulations of the first stars \citep{abel2002} and earliest galaxies \citep{wise2012}, simulations of Milky-Way analogue disk galaxies \citep{goldbaum2016}, and large-scale cosmological simulations \citep{xu2016}.

The \texttt{FLASH} code \citep{fryxell2000} is similarly intended as a community resource. Rather than in the \texttt{Enzo} model where the community develops the code, \texttt{FLASH} initially had funding for full-time developers and is to this day mainly developed in private at the University of Chicago, although community contributions are regularly integrated into \texttt{FLASH} releases \citep{dubey2013}. Rather than being publicly available, access to the source code must be requested via a web form at the \texttt{FLASH} website (\url{http://flash.uchicago.edu}) after agreeing to cite the authors in any publication making use of the code. \texttt{FLASH} is a modular fortran codebase and includes a variety of problem types, hydrodynamics, MHD, and gravity solvers, as well as modules for nuclear burning and multispecies equations of state out of the box that can be used for a wide variety of applications. Applications of \texttt{FLASH} in the literature include simulations of interstellar turbulence \citep{federrath2013}, simulations of supernova progenitors \citep{couch2014}, and tidal disruption events \citep{guillochon2013}.

Within the numerical relativity community the \texttt{Einstein Toolkit} is the tool of choice for numerically solving the equations of general relatavistic magnetohydrodynamics (GRMHD) and Einstein's equations \citep{loffler2012}. The \texttt{Einstein Toolkit} is based on the \texttt{Cactus} framework and the \texttt{Carpet} AMR driver and is developed in the open as a community resource under the GPL \citep{loffler2013} ({\small \url{https://bitbucket.org/einsteintoolkit/einsteintoolkit}}) while the \texttt{Cactus} framework is distributed under the LGPL.\@  The most common scientific use case for the \texttt{Einstein Toolkit} is numerical simulations of compact objects, including supernova progenitors \citep{mosta2014} and particular compact object mergers in support of grtavitational wave observations \citep{ajith2012}. Numerical simulations of these events are used to generate model gravitational wave signals that are fit to observations from LIGO to infer properties of the systems that emanate gravitational wave signals such as mass ratio and orbital parameters.

Other open astrophysics simulation codes making use of AMR include \texttt{RAMSES} \citep{teyssier2002} and \texttt{Athena++}, the latest version of the \texttt{Athena} simulation code \citep{stone2008}. \texttt{RAMSES} transitioned to open development on Bitbucket relatively recently at \url{https://bitbucket.org/rteyssie/ramses}, while \texttt{Athena} is still a private code with a public version including limited support for advanced physics modules, available at \url{http://princetonuniversity.github.io/athena/}.

\section{New Codes and Future Directions}
\label{usm}

The AMR and SPH codes described above have had much success. However, over the past decade we have seen some movement away from these codes in favor of new computational techniques and codes that are able to exploit new supercomputer architectures. Examples of these new codes are briefly described here.

Moving mesh methods can be thought of as a synthesis of AMR and SPH.\@  In this approach, the equations of hydrodynamics are solved on a voronoi mesh generated by a set of Lagrangian points. Like SPH particles, these points follow the fluid flow and naturally end up located in regions that need enhanced resolution. Like an AMR code, the equations of fluid dynamics are solved using finite volume mthods and are evaluated using similar techniques to the PPM method described above. By far the most prolific moving mesh code in terms of published scientific results is the \texttt{AREPO} code \citep{springel2010}. \texttt{AREPO} can be thought of as an evolution of the \texttt{Gadget} SPH code, making use of the same data format and graivtational force algorithm, but with a completely updated and reworked hydrodynamics module. \texttt{AREPO} is a private code, although it has already been used extensively in the literature. Science results produced using moving mesh codes include the Illustris simulations \citep{vogelsberger2014}, a suite of large-scale N-body/hydrodynamics cosmology simulations, as well as simulations of other astrophysics phenomena such as AGN jets \citep{weinberger2017}. Another moving mesh code is \texttt{Shadowfax} \citep{vandenbroucke2016} (\url{http://www.dwarfs.ugent.be/shadowfax/}), publicly available under the GPLv3.

The \texttt{GIZMO} code \citep{hopkins2015} is also an evolutionary descendent of the \texttt{Gadget} code that makes use of an improved hydrodynamics formulation. Gizmo is not a moving mesh code. Instead, its hydrodynamics module can be thought of a a novel reformaulation of meshless hydrodynamcis that is fundamentally different from smoothed particle hydrodynamics. Compared with SPH, it exhibits improved capturing of shocks, instabilities, and mixing. The major science results produced using Gizmo come from the FIRE simulation suite \citep{hopkins2017}, which captures the formation of a Milky Way analogue disk galaxy in a cosmological context using the zoom-in technique. There is a public release of \texttt{GIZMO} under the GPLv2 license ({\small \url{http://www.tapir.caltech.edu/~phopkins/Site/GIZMO.html}}) but the public release does include the full complement of physics packages available in the private version of Gizmo. According to the \texttt{GIZMO} website, access to the private version that includes a full complement of physics modules is available upon request.


Finally, I would like to finish by highlighting the \texttt{GAMER} code \citep{schive2010}, a block-structured AMR code written to maximize the efficiency of modern GPU-powered supercomputer architectures. \texttt{GAMER} includes a number of physics modules that make it suitable for a large number of astrophysics applications, including N-body particle dynamics, MHD, star formation and feedback, and chemistry. Due to its use of GPU accelerators, \texttt{GAMER} is also substantially faster than CPU-bound AMR codes like \texttt{FLASH} and \texttt{Enzo}. \texttt{GAMER} will be publicly available in the near future.

\acknowledgements\ Nathan Goldbaum acknowledges support via NSF grant ACI-1535651 and via the Gordon and Betty Moore Foundation's Data-Driven Discovery Initiative through Grant GBMF4651.

\bibliography{I5-1.bib}

\end{document}